\documentclass[pre,twocolumn,showpacs,preprintnumbers,amsmath,amssymb]{revtex4}
\usepackage{simplewick}

\usepackage{graphicx}
\usepackage{dcolumn}
\usepackage{bm}
\newcommand \be{\begin{eqnarray}}
\newcommand \ee{\end{eqnarray}}
\newcommand \ba{\begin{align}}
\newcommand \eea{\end{align}}

\begin{document}

\title
{Dynamical mechanism of antifreeze proteins to prevent ice growth}
\author
{B. Kutschan$^1$, K. Morawetz$^{1,2,3}$, S. Thoms$^4$
}

\affiliation{
$^1$ M\"unster University of Applied Science, Stegerwaldstrasse 39, 48565 Steinfurt, Germany}
\affiliation{
$^2$International Institute of Physics (IIP)
Federal University of Rio Grande do Norte
Av. Odilon Gomes de Lima 1722, 59078-400 Natal, Brazil}
\affiliation{
$^3$Max-Planck-Institute for the Physics of Complex Systems,
  01187 Dresden, Germany}
\affiliation{
$^4$Alfred Wegener Institut, Am Handelshafen 12, D-27570 Bremerhaven, Germany}


\begin{abstract}
The fascinating ability of algae, insects and fishes to survive at temperatures
below normal freezing is realized by antifreeze proteins (AFPs).  
These are surface-active molecules and interact with the diffusive water/ice interface thus
preventing complete solidification. We propose a new dynamical mechanism on how these proteins inhibit the freezing of water. We apply a Ginzburg-Landau type approach to describe the
phase separation in the two-component system (ice, AFP). The free energy density involves two fields: one for the ice phase with a low AFP concentration, and one for liquid water with a high AFP concentration. The time evolution of the ice reveals microstructures resulting from phase
separation in the presence of AFPs. We observed a faster clustering of pre-ice structure connected to a
locking of grain size by the action of AFP, which is an essentially dynamical
process. The
adsorption of additional
water molecules is inhibited and the further growth of ice grains stopped. 
The interfacial energy between ice and water is lowered allowing the AFPs
to form smaller critical ice nuclei. Similar to a hysteresis in magnetic materials we observe a thermodynamic hysteresis leading to a nonlinear density dependence of the freezing point depression in agreement with the experiments. 
\end{abstract}
\maketitle

\section{Introduction}

\begin{figure}[h]
\centerline{
\includegraphics[width=2cm]{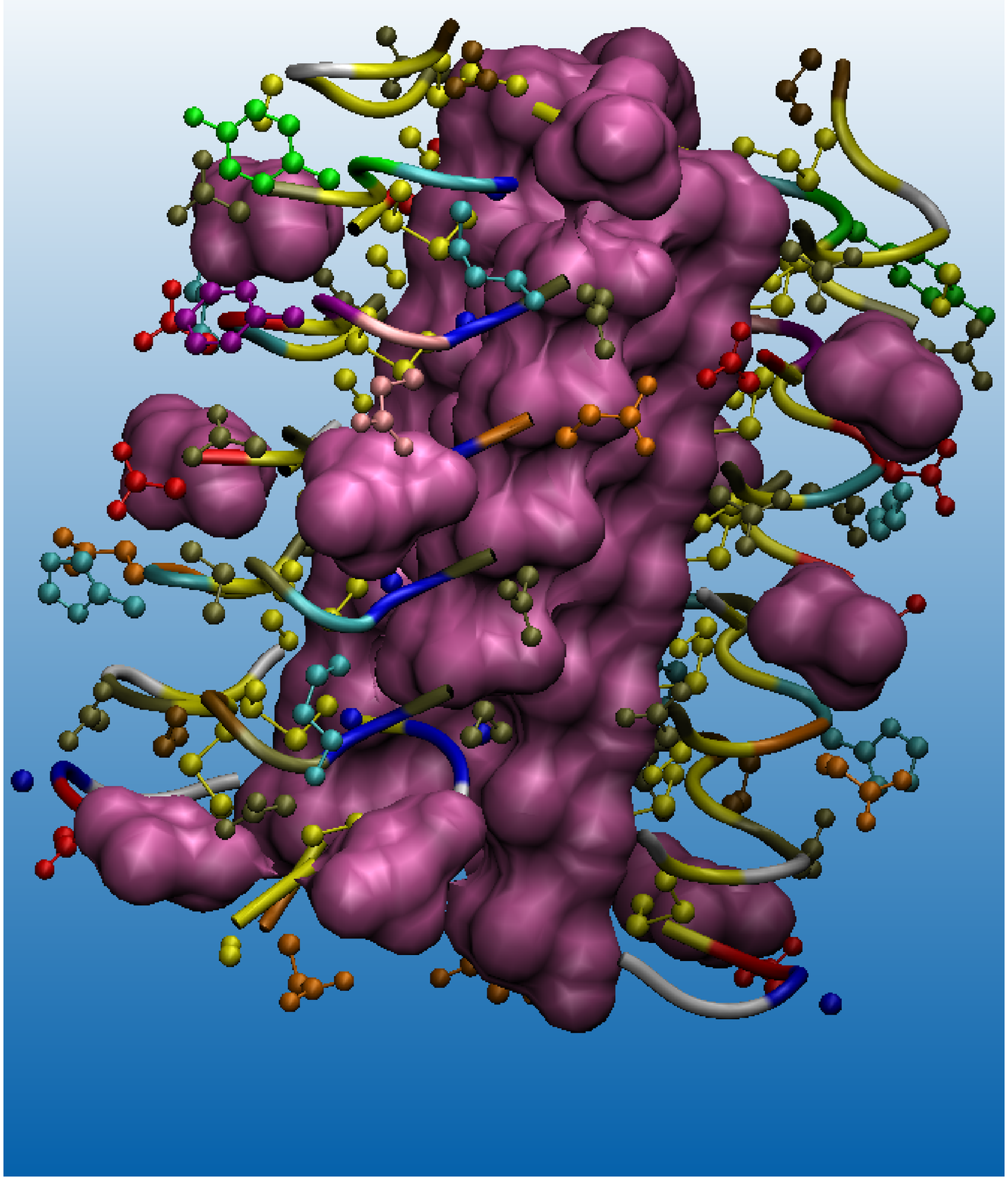}
\includegraphics[width=2cm]{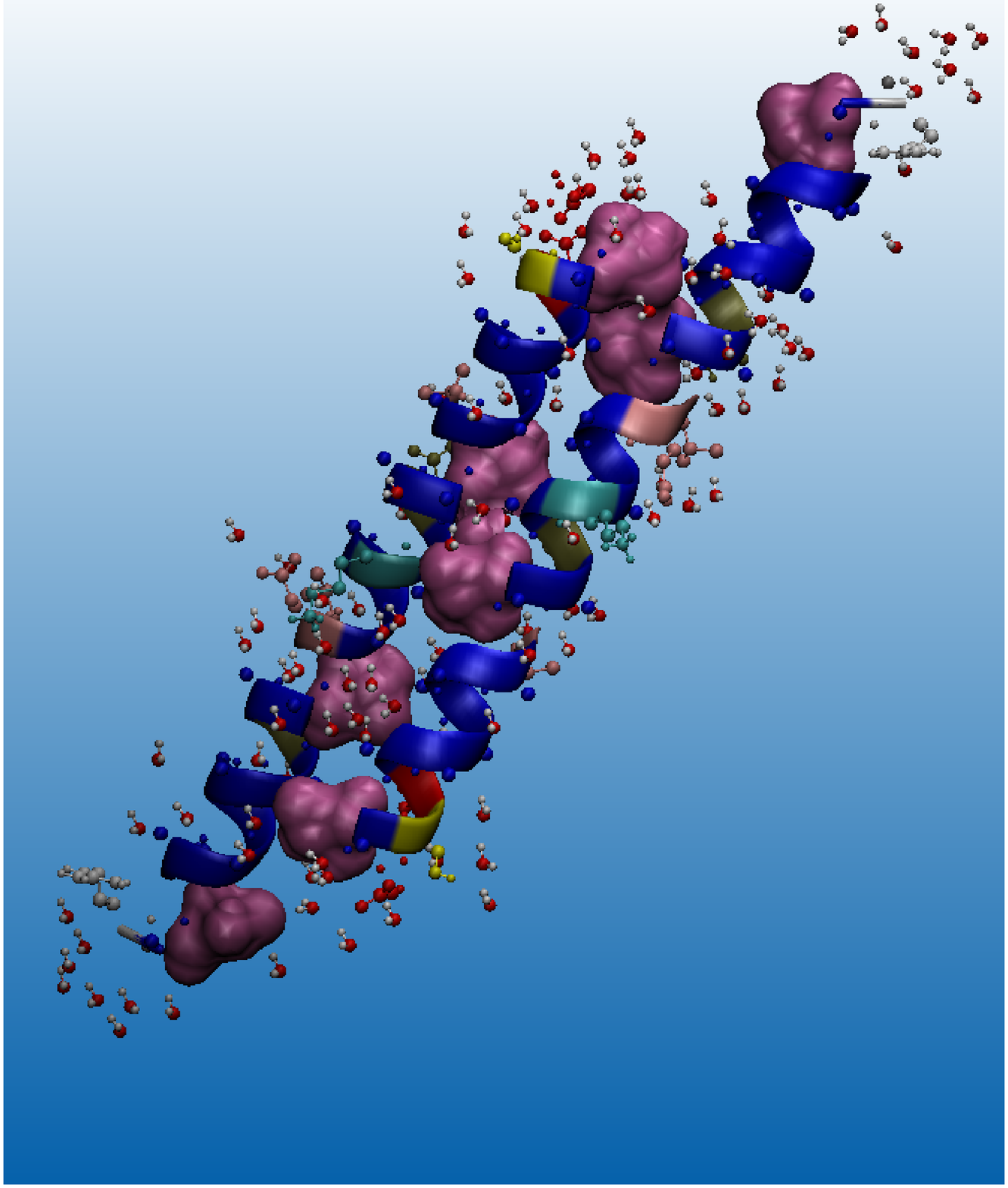}
\includegraphics[width=2cm]{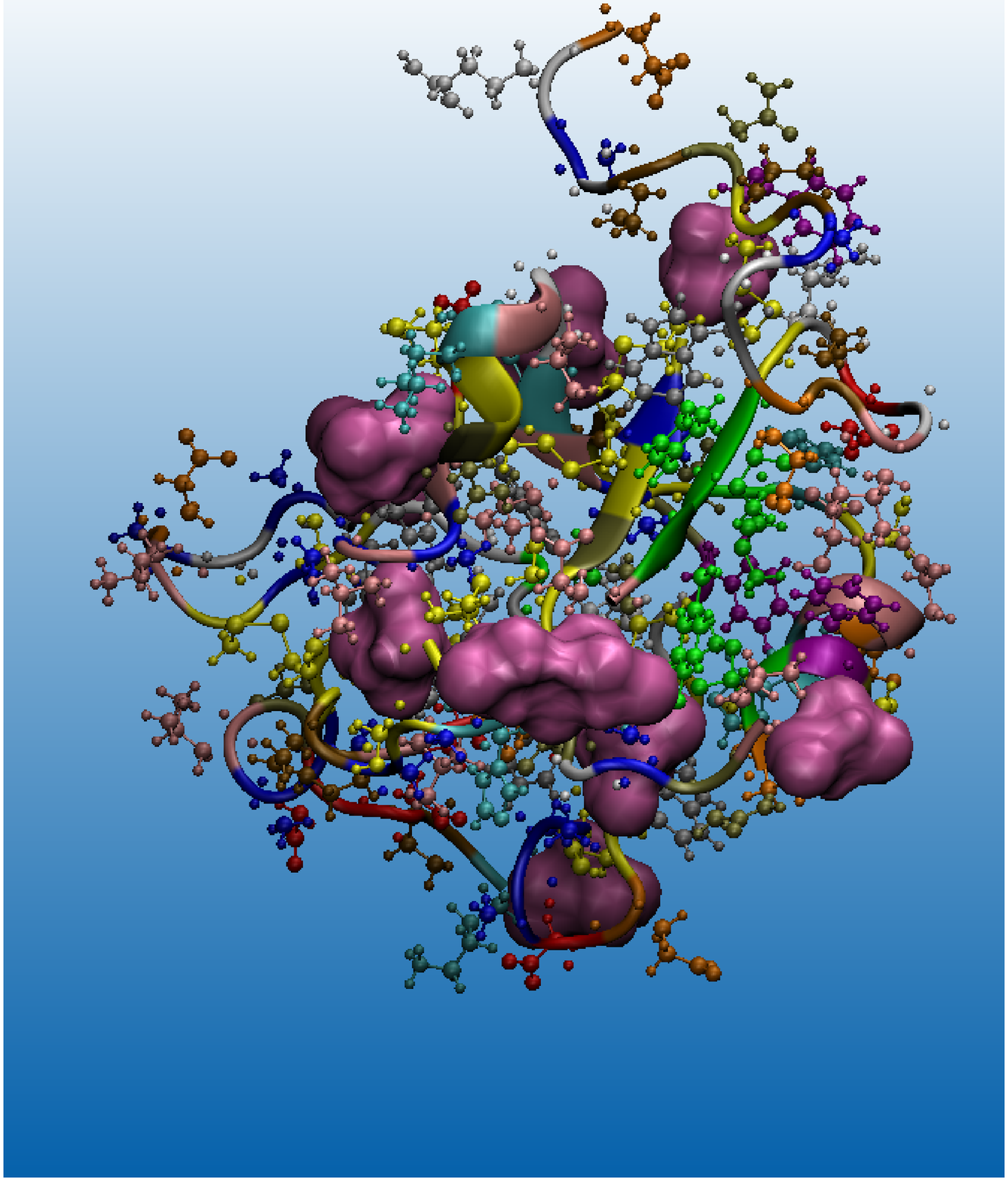}
\includegraphics[width=2cm]{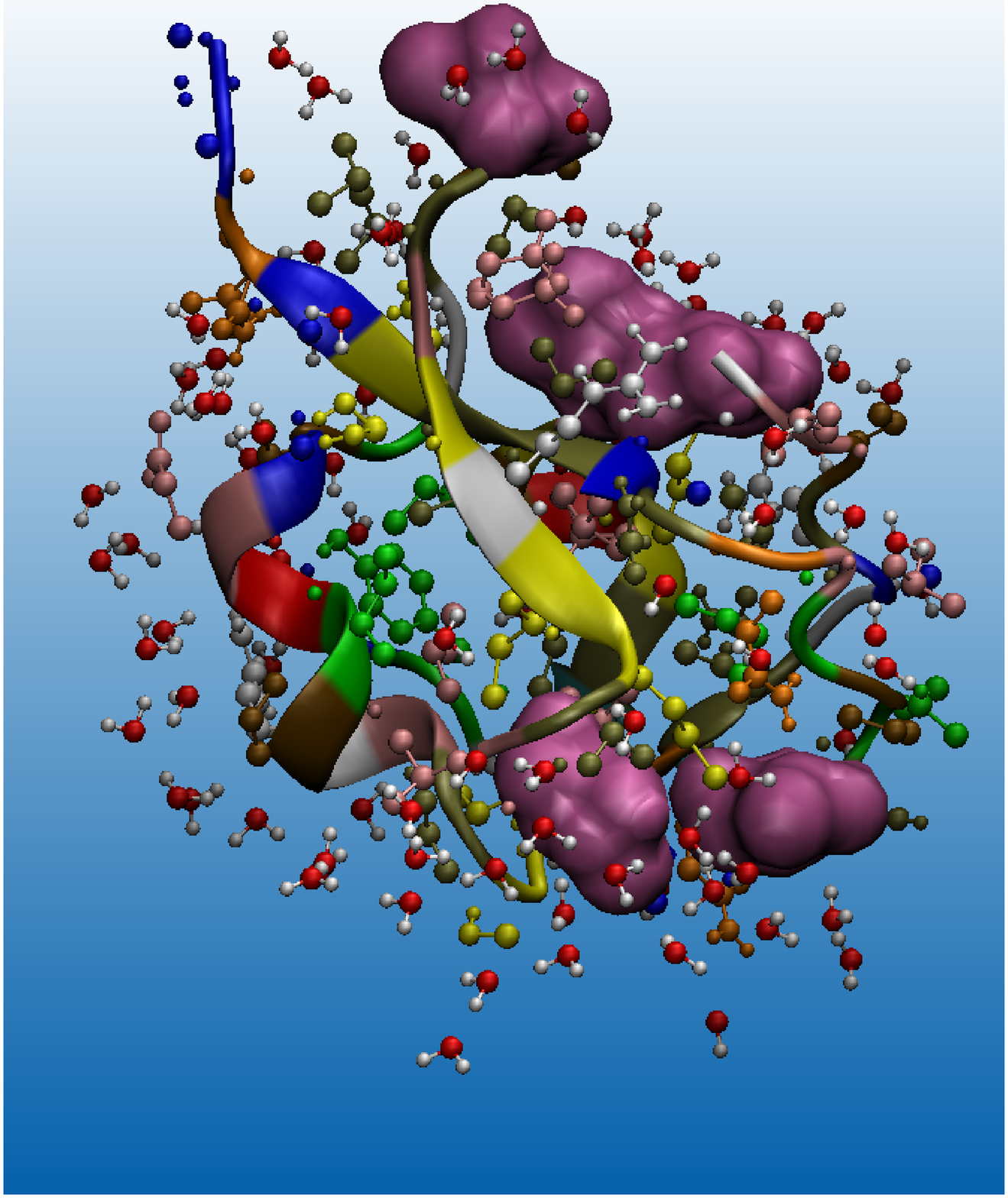}
}
\caption{(Color online) Four different classes of considered AFP structures (from left to right):  
tenebrio
molitor (1EZG) as used in \cite{mao}, psodeupleuronectes americanus (1WFB) in \cite{HSGDF11,SH81},
hemitripterus americanus (2AFP) in \cite{GLLDSDS98},  and macrozoarces americanus (1MSI) in \cite{WAW12}. Figures prepared with crystallographic data in RCSB Protein Data Bank. 
\label{struct}}
\end{figure}

The suppression of freezing temperature by antifreeze proteins (AFPs), which allows fishes, plants and diatoms to survive even below $ 0\,^{\circ}\mathrm{C} $ is a fascinating phenomenon. Their activity was first observed in Arctic fishes in 1957 \cite{Sch57}. AFPs were isolated in 1969 \cite{DeVries} and were discovered in plants in 1992 \cite{GAYHM92,UDK92}. For an overview see \cite{HAH03}. Four classes (I-IV) of antifreeze proteins are known, with wide structural diversity and sizes. These include a class of antifreeze glycoproteins
(AFGPs) and a number of hyperactive antifreeze proteins in insects
\cite{Ba01,Jia} as illustrated in figure~\ref{struct}. A large diversity of molecular structures is apparent and consequently their influence on 
antifreeze activity and grain growth \cite{yeh1994} has to be considered.
Due to the multiple hydrophilic ice-binding domains, the AFPs inhibit ice recrystallization and nucleation \cite{GriYa04}. The difference between the temperature below which the AFPs cannot stop ice nucleation and the melting temperature of ice crystals in solution is called thermal hysteresis.
Experimental results illustrate a connection between protein structure and the
thermal hysteresis activity on the one hand \cite{Bu86,NF12}, and between the
protein structure and the ice growth patterns on the other
\cite{BaCe12,yeh1994}.

The mechanism of AFP binding is still unresolved
\cite{Hew} since the details of the antifreeze effect are difficult to test
experimentally, mainly because it is not easy to access them in their natural environment. Molecular dynamic simulations \cite{NF12,Liu2012} are limited by the computing power and running times. Moreover the interaction between AFPs and the liquid-solid interface is determined by the choice of the water model and the potential parameters \cite{Guillot,Berendsen,BePoGu,Svishchev,Jo,Kiss,Stern,Mahoney,Ess}. The simple freezing point depression such as observed in saline solutions is proportional to the molal concentration of the solute molecule
\cite{YF96}, where the colligative property does not depend on the structure of the
molecules. However, a more careful inspection of AFPs shows a nonlinear dependence so that colligative effects are ruled out \cite{NF12}. 
We propose a dynamical mechanism for freezing point depression by AFPs
leading to a nonlinear dependence on the AFP density.

There is not a single mechanism known up to date which explains the AFP ice-binding affinity and specificity \cite{Ba01}. One reason lies in the
considerable variation of the primary, secondary and tertiary structure of
AFPs \cite{Ba01,Jia,YF96,GruRi09,HaWa99,Hew} because the ice-binding affinity
depends on the molecular recognition according to the key-lock principle
\cite{Daley,Leinala}. The different kinds of structures lead to various
kinetic models of AFP activity
\cite{LYLFK06,LLL09,Bu86,KrZa05,Kubota,WAW12,sander}. A significant difference
can be found by the shape of the antifreeze activity between AFGP 4 and AFGP 8
\cite{LiLuo,Bu86}. In addition to that, several authors have substituted threonine
residues and have investigated the influence on the thermal hysteresis
\cite{Zhang,Schrag} or have  enhanced the activity of a $\beta$-helical antifreeze
protein by engineered addition of coils \cite{marshall,mao}. 

Moreover, other
authors have assessed the relationship between the non-colligative freezing point depression and
the molecular weight of AFPs \cite{Schrag,raymond} and have observed cooperative
functioning between the larger and the smaller components \cite{Osuga}. 
The stabilization of supercooled fluids by thermal hysteresis proteins
\cite{Wilson,Zachariassen01} is discussed with
a number of models based on kinetics, statistical mechanics and homogeneous or
inhomogeneous nucleation
\cite{LYLFK06,LLL09,LiLu,Kubota,LiLu1994,LiLuo,LL03,LL06,WAW12,sander,mao,raymond}. 
All these models describe a thermal hysteresis as a function
of concentration, but no pattern formation in space and time during the phase
transition from liquid water to ice. In contrast, the Turing model \cite{KMG09} and the phase field model \cite{Thoms,MTK14} simulate the morphology of the microstructural growth but without thermal hysteresis.
Our main goal is to include kinetic models 
capable to describe the experimental observations
into the phase field method
 and to justify the non-equilibrium stabilized supercooled
liquid state of the hysteresis.

We use coupled phase field
equations for the water-ice structure and the AFP concentration, which is 
connected with the question on how the protein structure influences antifreeze 
activity. 
The various phase field approaches differ in the choice of the bulk free energy 
density. We use the Gibbs free energy according to the classical Landau theory 
of a first-order phase transition as the freezing of water. This 
phenomenological free energy as a form of mean-field theory is expanded in a power series of
the order parameter. The Landau coefficients describe the equilibrium as well as the metastable states of the ice/water system and can be deduced from properties of water as demonstrated in \cite{Thoms}. A more microscopic approach has to rely on water models like density functionals, which are currently not simultaneously available for water and ice. We therefore restrict ourselves to three phenomenological parameters, which can be linked to known water properties.

The dynamical formation of the micro-structures are calculated by
solving the coupled phase field equations which combine the phase field theory of Caginalp \cite{Caginalp_01,Caginalp_87,Caginalp_88,Caginalp_89,Caginalp_90,Caginalp_91}, Cahn and Hilliard \cite{Gr2,Gr1,Gr3,Gr4} with various kinetics \cite{YF96,Bu86,Kubota,WAW12,mao}. In the next section we outline shortly the basic equations and models. Then we present the nonlinear freezing point depression from static conditions and support this by numerical discussions of the time-dependence of freezing suppression.

\section{Coupling of AFP to ice structure}

We begin with a phase
field model \cite{Gr2} with ice nucleation of Cahn-Hilliard-type proposed in \cite{Gr1}, due to the lack of complete
understanding of water by first principles.
We identify the ice structure by an order parameter $u$ describing the ''tetrahedricity'' \cite{Me}
\begin{equation}
u \sim 1-M_T = 1- \frac{1}{15 <l^2>}\sum_{i,j}(l_i - l_j)^2 ,
\end{equation}  
where $l_i$ are the lengths of the six edges of the tetrahedron
formed by the four nearest neighbors of the considered water molecule. For an
ideal tetrahedron one has $M_T=0$ and the random structure yields
$M_T=1$. In this way it is possible to discriminate between ice- and
water molecules either by using a two-state continuous function. Other authors prefer the ''tetrahedrality'' in order to define the order parameter \cite{Err2001,Kumar2009,Mason2007}. 
We adopted a quartic order parameter relationship 
of Ginzburg-Landau-type for the free energy density
\begin{equation}
f(u,v)\sim \beta u + \lambda u^2 - 2\lambda u^3 + \lambda u^4+ c\left(\frac{\partial u}{\partial x}\right)^2
\label{set1}
\end{equation}
allowing nonlinear structures to be formed. The
fourth order function enables a double well potential for the description of
the water-ice phase transition \cite{HaOx87}.  
The coefficient $\lambda$ describes the free
energy density scale and $\beta$ the deviation from equilibrium.

The versatile action of different molecular structures of AFPs on the grain
growths is simulated by an
activity parameter relating the structural order parameter to the antifreeze
concentration
$
f(u,v)\sim - a_1uv 
$.
The coefficient $a_1$ describes the interaction between the order parameter $u$ and the
antifreeze concentration $v$. This approach is different from the description
of simple freezing point depression in saline solutions \cite{Thoms,MTK14} in that
the AFPs act analogously to a magnetic field on charged particles thus providing a hysteresis.

We follow the philosophy of the conserving Cahn-Hilliard equation 
and assume that the time-change of the order parameter is proportional to the gradient of a current $\dot u=-\nabla j$. The current itself is assumed to be a generalized force, which in turn is given by the gradient of a potential, $j\sim \nabla \Phi$. The latter is the variation of the free energy such that finally $\dot u=-\nabla^2(-{\partial f\over \partial u})$ results. From the differential form for a general continuity equation $\frac{\partial v}{\partial t} + \frac{\partial j}{\partial x} = 0 $ with the AFP field $v$ and the flux $j = -\frac{\partial}{\partial x}\left(a_3v+a_2u\right)$ we get a diffusion equation
$\frac{\partial v}{\partial t}  = \frac{\partial^2}{\partial x^2}\left(a_2u + a_3v \right) \label{ev1}$ as evolution for the AFP concentration. 
For  further considerations  we introduce the dimensionless quantities $\tau=\frac{t}{t_0}$, $\xi=\frac{x}{x_0}$, $u=C_1\hat\psi$, $v=C_2\rho$,  with $C_1=1$, $C_2=\frac{a_1}{12\lambda}$, $x_0=\frac{1}{6}\sqrt{\frac{6c}{\lambda}}$, $t_0=\frac{c}{72\lambda^2}$ with $\alpha_1=\frac{a^2_1}{12^2\lambda^2}$,  $\alpha_2=\frac{a_2}{12\lambda}$ and $\alpha_3=\frac{c\beta}{72\lambda^2}$ such that the order-parameter and the AFP concentration equations read 
\begin{eqnarray}
\frac{\partial\psi}{\partial \tau}  \!&=&\! \frac{\partial^2}{\partial \xi^2}
\!\left (\!
\alpha_3\!+\!\left(\frac{1}{6}\!-\!a\!+\!a^2\right) \psi 
\!+\!\left(a\!-\!\frac{1}{2} \right) \psi^2 \!+\!\frac{1}{3}\psi^3 
\right .
\nonumber\\
&&\left . -\alpha_1\rho - \frac{\partial^2 \psi}{\partial \xi^2} 
\right ) 
\label{ff6c}\\ 
\frac{\partial \rho}{\partial \tau}  &=& \frac{\partial^2}{\partial \xi^2}\left(\psi + \alpha_2 \rho \right) .
\label{ff6d}
\end{eqnarray}
We have shifted $\psi=\hat\psi - a$ by technical reasons since $a=1/2$ allows the removal of the cubic term in the free energy.

The Cahn-Hilliard equation 
cannot be derived from the Onsager reciprocal theorem \cite{Whe} because the Markovian condition is not satisfied \cite{Emm},
but has the virtue of a conserving quantity. It possesses a stationary solution, which can be seen from the time evolution of the total free energy \cite{Beste}
\be
{d\over d t} \int f dV&=&\int {\delta f\over \delta \psi} {d \psi\over d t} dV=\int {\partial f\over \partial \psi}\nabla^2({\partial f\over \partial \psi})dV
\nonumber\\&=&-\int \left (\nabla {\partial f\over \partial \psi}\right )^2 dV\le 0.
\ee

\section{Mechanism of freezing point depression}
\subsection{Phase diagram}

We analyze the static solution of (\ref{ff6c}) and (\ref{ff6d}) eliminating $\rho=-\psi/\alpha_2$ and obtain
\begin{equation}
\frac{\partial^2 \psi}{\partial \xi^2}=\alpha_3-\aleph\psi - \left(a-\frac{1}{2}\right)\psi^2 +\frac{1}{3}\psi^3 \equiv \frac{\partial \phi}{\partial \psi}\label{f11}
\end{equation} 
which can be understood as the Euler-Lagrange equations from the free energy 
\ba
f&=\alpha_3\psi-\aleph {\psi^2\over 2} +\left(a-\frac{1}{2} \right) {\psi^3\over 3} +\frac{1}{12}{\psi^4} + \frac 1 2 \left (\frac{\partial\psi}{\partial \xi}\right )^2
\nonumber\\&\equiv\phi+\frac 1 2 \left ({\partial \psi\over\partial \xi}\right )^2
\end{align}
where we abbreviate $\aleph = - \frac{1}{6} +a -a^2-\frac{\alpha_1}{\alpha_2}$. 
Eq. (\ref{f11}) is easily integrate by multiplying both sides with $\partial \psi/\partial\xi$ to yield 
\begin{equation}
\frac{1}{2}\left(\frac{\partial \psi}{\partial \xi}\right)^2 = \phi+c\equiv\Phi \label{virial}
\end{equation}
providing the virial theorem $f=2\phi+c=2\Phi-c=(\psi')^2-c$ as an expression for the conservation of energy. This fact allows us to calculate the static profile of the kink solution as shown.
Depending on the parameter $\alpha_1/\alpha_2$ and $\alpha_3$ we obtain
an asymmetric potential $\Phi$ describing the thermodynamic hysteresis though
$\alpha_3$ does not influence the
dynamics due to the differentiation in Eq. (\ref{ff6c}). Near the phase transition the potential $\Phi$ becomes symmetric and
possesses a reflection symmetry $\Phi(\psi)=\Phi(-\psi)$ with the minimum
$\psi_{min}=\pm\sqrt{3\aleph}$. Hence, the constant $c$ can be found from the condition $\Phi(\sqrt{3\aleph})=-\frac{3}{4}\aleph^2+c=0$.

\begin{figure}[h]
\centering
\includegraphics[width=8.5cm
]{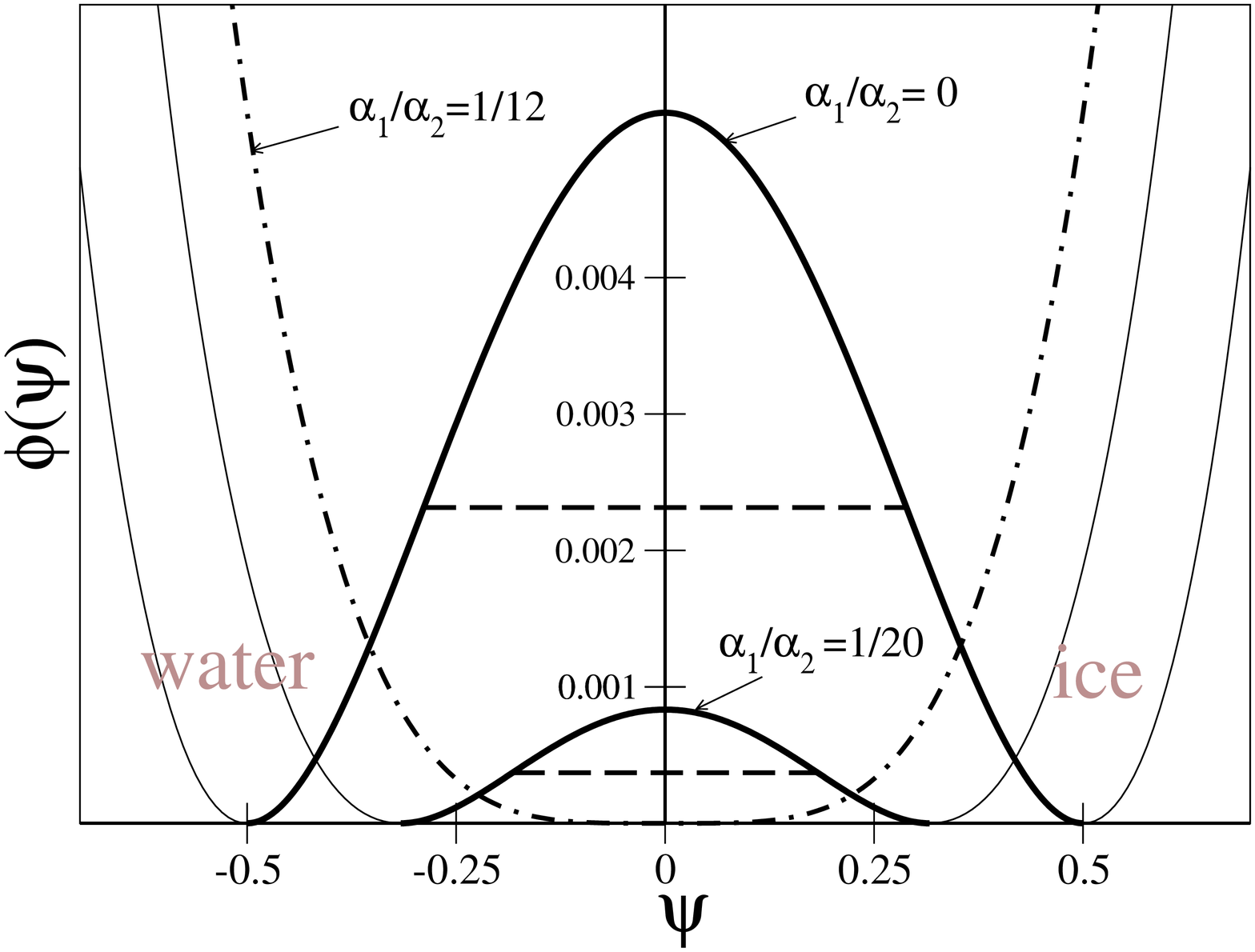}
\caption{Free energy density (solid) versus order parameter for different AFP concentrations, pure ice/water $\frac{\alpha_1}{\alpha_2}=0$ (upper line). The freezing transition interval is given by $\partial^2 \phi/\partial\psi^2=0$ (dashed lines). \label{potential_02}}
\end{figure}

In figure \ref{potential_02} the symmetric potential is plotted where the left
and right minima reflect the stable phase of water and ice respectively. The
concave $\partial^2\Phi/\partial\psi^2 <0$ region corresponds to a negative
diffusion coefficient leading to structure formation. The flux diffuses up against the concentration gradient contrary to the Onsager reciprocal theorem \cite{Whe} and which is the unstable phase transition region. This freezing region is reduced by the AFP concentration where for $\frac{\alpha_1}{\alpha_2}=\frac{1}{12}$ the double well vanishes.

\subsection{Linear stability analysis}

\begin{figure}[h]
\centering
\includegraphics[width=8.5cm
]{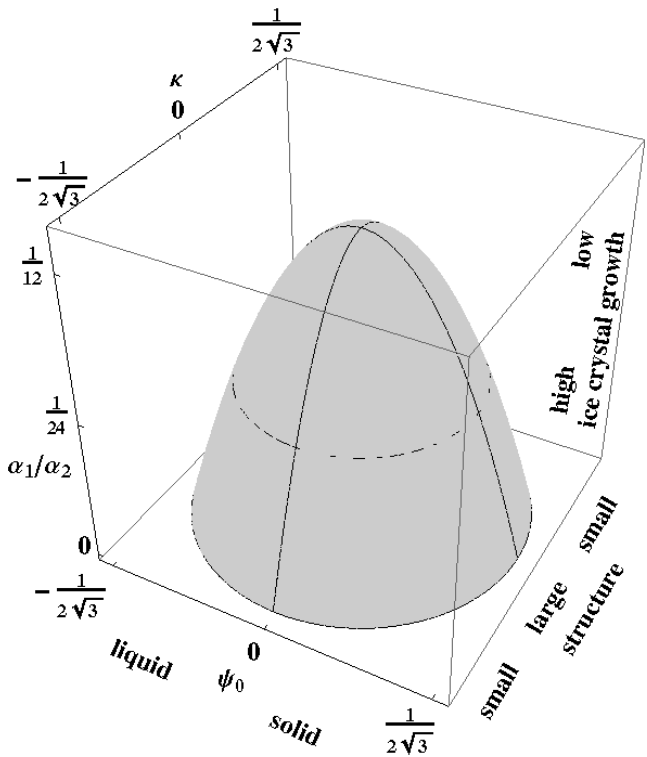}
\caption{Freezing (spinodal) region dependent on the order parameter $\psi_0$ and the wavenumbers $\kappa$ and the thermal hysteresis activity $\frac{\alpha_1}{\alpha_2}$. \label{z_psi}}
\end{figure}

It is also possible to compute the phase transition region with the help of the positive eigenvalues $\mu$ of the linear stability analysis 
around the equilibrium value $\psi=\psi_0+\psi_0\exp(\mu \tau + i \kappa \xi)$
and $\rho=\rho_0+\rho_0\exp(\mu \tau + i \kappa \xi)$ with the wavenumber
$\kappa$ and the fixed point $\psi_0$.  Each fixed point describes a spatial
homogeneous order parameter $\psi=\psi_0=const$ and corresponds to a
stationary solution of water or ice. For pure water ($\alpha_1=0$) of
(\ref{ff6c}) the range of unstable homogeneous solution is 
\begin{equation}
\mu^*(\kappa)=\left(\aleph-\psi^2_0\right)\kappa^2-\kappa^4>0 \label{eig}
\end{equation}
and a perturbation grows in time and therefore establishes a spatial
structure. For the complete equation system (\ref{ff6c}) and (\ref{ff6d}) with thermal
hysteresis proteins (AFPs), one obtains the eigenvalues 
\ba
\mu(\kappa) = -\frac{1}{2}(\alpha_2 \kappa^2\!-\!\mu^*(\kappa)) \!\pm\! \frac{1}{2}\sqrt{(\alpha_2 \kappa^2 \!+\! \mu^*(\kappa))^2 \!-\! 4 \alpha_1 \kappa^4}.
\end{align}
The region of positive eigenvalues corresponds to the freezing (spinodal) 
region. The unstable modes vanish for $\psi^2_0>{1/12}$ and also the double well for ${\alpha_1/\alpha_2}>{1/12}$. A phase transition occurs only, if the fixed points are located inside of $-{1/\sqrt{12}} < \psi_0 < {1/\sqrt{12}}$ being also inside the freezing (spinodal) interval in Fig. \ref{potential_02}.
Since $\sqrt{ \aleph}=\psi_0$  at the  inflection points it can be concluded from Eq. (\ref{eig})
\begin{equation}
\kappa^2+ \psi_0^2-\left(\frac{1}{12}-\frac{\alpha_1}{\alpha_2}\right)\psi_0 <0
\end{equation}
forming an elliptic paraboloid as phase diagram in figure \ref{z_psi}. The
size of the microstructure is coupled to the AFP concentration and to the order parameter $\psi_0$ which
decides how much ice and water is present. The freezing region shrinks with
increasing AFP concentration and vanishes at $\alpha_1/\alpha_2=1/12$.

%
%
%
%
%
%
%
%

\subsection{Surface energy depression}
\begin{figure}[h]
\includegraphics[width=8cm]{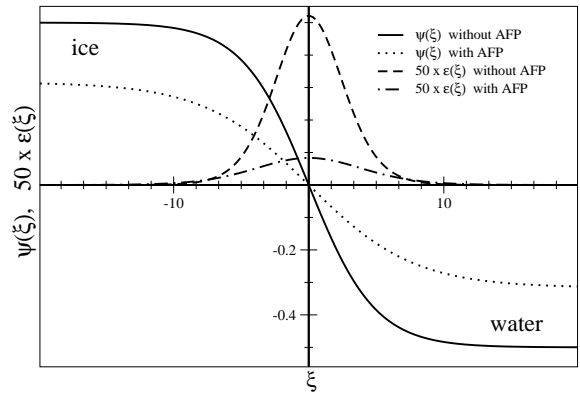}
\caption{Static kink solution (\ref{kink}) and the interfacial energy density $\varepsilon(\xi)$ on the interface with AFPs ($\frac{\alpha_1}{\alpha_2}=0.05$) and without AFPs ($\frac{\alpha_1}{\alpha_2}=0$).
\label{fig_intenergy}
}
\end{figure}

The virtue of the Cahn Hilliard equation (\ref{ff6c}) is the existence of a transient stationary solution by integrating (\ref{virial}) in form of a kink solution
\begin{equation}
\psi(\xi)=-\sqrt{3 \aleph}\tanh\left[\sqrt{\aleph/2}(\xi-\xi_0)\right] \label{kink}
\end{equation} 
with $\aleph=\frac{1}{12} -\frac{\alpha_1}{\alpha_2}$ as plotted in figure
\ref{fig_intenergy}.
The corresponding interfacial energy density of the kink is easily evaluated
using the centered free energy density
$\epsilon(\xi)=f(\xi)-c = ({\partial \psi \over \partial \xi})^2$
plotted as well in figure \ref{fig_intenergy}.

One recognizes that the presence of AFPs reduces the kink between water and ice and lowers the interfacial energy density. This is already a static mechanism which shows how the AFPs inhibit the formation of large ice clusters. The interfacial surface energy (tension) 
\begin{equation}
\zeta = \int\limits^\infty_{-\infty} \varepsilon(\xi)d\xi =  (2 \aleph)^{3/2}
\label{zeta}
\end{equation}
decreases with increasing AFP coupling and finally vanishes at the critical value
$\alpha_1/\alpha_2=$1/12 which is the limit of the stability region.
Transforming back to the dimensional interfacial
energy one obtains  $6^{3/2}\gamma\zeta$ with our choice of the surface tension $\gamma=21.9\frac{mJ}{m^2}$ \cite{Gr1}. Direct and indirect measurements provide values which vary between $20\frac{mJ}{m^2}$ and $46\frac{mJ}{m^2}$ \cite{Jo1974}. In contrast to antifreeze proteins, $\gamma$ increases
linearly with the salt
concentration and a larger
critical nucleus is required in order to generate an interface compared to pure
water. Salt inhibits the nucleation process because of the
higher energy threshold whereas thermal hysteresis proteins
(AFPs) reduce the threshold for a stable nucleus. From this effective 
surface tension one might fix the ratio of materials parameter 
$\aleph=1/12-\alpha_1/\alpha_2$.

An ice crystal forms when the aqueous phase undergoes supercooling below the freezing point. In the supercooled phase water can transform from the aqueous to the solid phase by the growth of nucleation kernels if a critical size is exceeded. This critical radius separates the reversible accumulation of ice molds (embryos) from the irreversible growth of ice crystal. The nucleation process happens in two steps. As soon as the nucleation embryo overcomes the critical size it starts growing irreversibly. This is a dynamical process which we will consider below. One can estimate the critical radius within a simple liquid drop model. The volume part of Gibb's potential is negative $\sim-{4 \pi r^3} \Delta G_V/3$ while the surface part contributes positively $\sim 4 \pi r^2 \zeta$. Therefore Gibb's potential exhibits a maximum at the critical cluster size 
$
r^{\ast} ={2 \zeta}/{\Delta G_V}.
$
As long as $r < r^{\ast}$ nucleation might happen (embryo) but no cluster grows for which case the critical cluster size has to be exceeded. This interfacial energy of the critical nucleus and the degree of supercooling are essentially influenced by the AFPs. Indeed the change of the free energy
between ice and water $\psi(\pm \infty)=\pm \sqrt{3 \aleph}$ reads with the stationary solution (\ref{kink})
\be
\Delta F=-2 (\alpha_3-\alpha_1 \rho) \sqrt{3 \aleph}\approx\Delta G_V.
\label{delta}
\ee 
With (\ref{zeta}) the critical (dimensionless) radius
$r^{\ast}={\sqrt{2\aleph} / 3 (\alpha_1 \rho-\alpha_3)}$
decreases as the AFP concentration $\alpha_1/\alpha_2$ increases. In other words more AFPs allow more ice nucleation but inhibit the cluster growth.

\begin{figure}[h]
\parbox[]{7.5cm}{
\includegraphics[width=7.5cm]{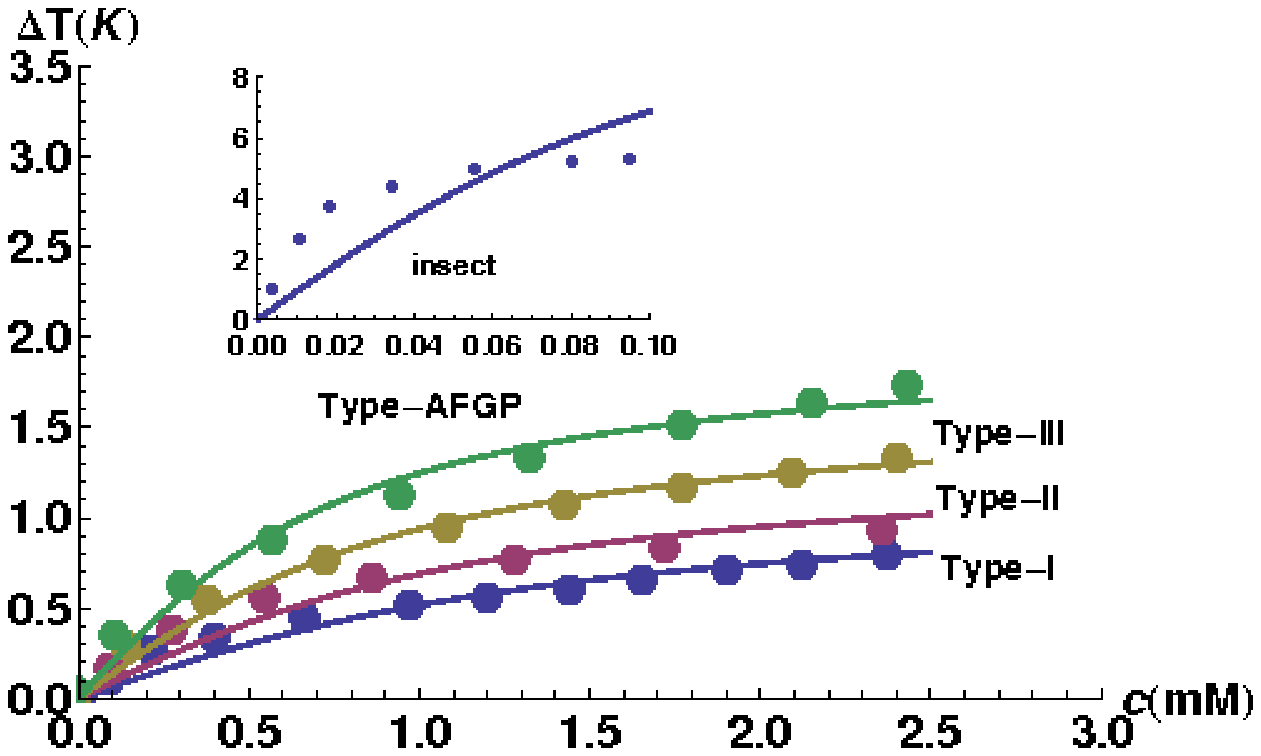}\\[2ex]
{
\begin{tabular}{c|ccccc}
Type & I & II & III & AFGP &insects\\
\hline
a&
 0.81 &
 1.08 &  
 2.58 &  
 4.31 &
34.39  
\\
\hline
b&
 0.44 &
0.37 &
0.88 &
1.13 &
0.05
\end{tabular}
}
}
\caption{(Color online) The freezing temperature depression of four different classes of AFP structures \cite{WAW12,GLLDSDS98,HSGDF11} and insects \cite{mao} versus AFP concentration where (\ref{tt}) is used (lines) to fit the experimental data (points) of the collected data in \cite{NF12}. The AFP-specific fitting parameter are given in the table.
\label{fig_comp}
}
\end{figure}

\subsection{Freezing point depression}

The decrease of the freezing temperature $\Delta T$ can be estimated by the
change of the free energy during the phase transition since the corresponding
minima of the asymmetric free energy are nearly the same as the ones of the
symmetric free energy (\ref{delta}). We can write
\be
\left . \Delta F\right |_{\rm ice-water}=\left .{\partial F\over \partial T}\right |_{\rm ice} \Delta T 
\label{dt}
\ee
for constant particle number and pressure. We model the temperature dependence of the coupling between AFP concentration and ice structure as
$
\alpha_1(T)=\alpha_0+\alpha_{10}(T-T_c^0)
$
with some internal threshold temperature $T_c^0$. The activity of AFP molecules will certainly be temperature dependent and cease to act at the critical temperature $\alpha_1(T^*)=0$. Therefore the freezing point depression is given by $\Delta T=T^*-T_c^0$. Introducing the AFP-dependent supercooling temperature $T_c=T_c^0-|\Delta T|$ one may write
$
\alpha_1(T)=\alpha_0 (T-T_c) / |\Delta T|
$.
From (\ref{dt}) and (\ref{delta}) we obtain the freezing point depression or
thermodynamical hysteresis observing $\psi|_{\rm ice}=\sqrt{3 \aleph}$ as
\be
|\Delta T|=\sqrt{\left ({b \over 2 \rho}\right )^2+a}-{b \over 2 \rho}
\label{tt}
\ee
with $a=2 \alpha_0 (T-T_c)/\alpha_{10}$ and $b=2 \alpha_3/\alpha_{10}$.
We see a nonlinear square root behavior of the freezing point depression which is dependent on the AFP concentration $\rho$  which expands for small concentrations
$
|\Delta T|\approx {a\over b}\rho= {\alpha_0 \over \alpha_3} (T-T_c) \rho
$
into a colligative freezing depression analogously as known from saline solutions.
The nonlinear behavior fits well with the experiments as seen in figure
\ref{fig_comp}. The fit parameters can be translated into two conditions
for the four materials parameter $\alpha_3, \alpha_{10}, \alpha_0$ and $T_c$
for each specific AFP. Together with the surface tension our approach leaves one
free parameter to describe further experimental constraints.

\begin{figure}[]
\includegraphics[width=6cm]{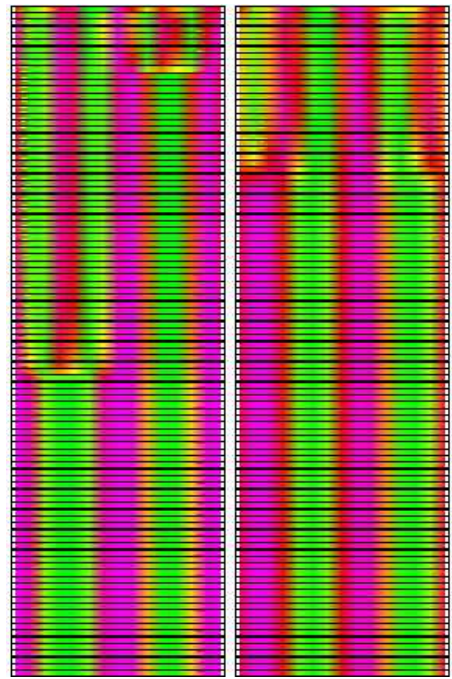}
\caption{
(Color online) Time evolution of the order parameter versus length from $1\times 10^5$
  to $2\times 10^6$ time steps (from above to below), left side without AFPs and right
  side with AFPs.
\label{afp_time}
}
\end{figure}

\subsection{Time evolution of ice growth inhibition}

Now that we have seen how the AFP reduces the interfacial energy and therefore the formation of ice crystals we turn to investigate the time evolution. We use the 
exponential time differencing scheme of second order (ETD2) \cite{CoMa}  with
the help of which a stiff differential equation of the type $y' = r y + z(y, t)$
with a linear term $r y$ and a nonlinear part $z(y, t)$ can be integrated. The
linear equation is solved formally and the integral over the nonlinear part is
approximated by a proper finite differencing scheme.
The time evolution 
of the coupled equations (\ref{ff6c}) are seen in figure \ref{afp_time}.
Due to the Cahn Hilliard equation we have conservation of total mass density of water $\rho_w$ but a relative redistribution between water $\rho_{liq}$ and ice $\rho_{ice}$ densities which read in terms of the ice order parameter $
\rho_w(\xi,t)=\rho_{liq}[1-\psi(\xi,t)]+\rho_{ice}[1+\psi(\xi,t)]$. 
Since the total integral over the order parameter $\int \psi(\xi,t)d \xi=const$ the mass conservation of water is ensured.

\begin{figure}[]
\includegraphics[width=8.5cm]{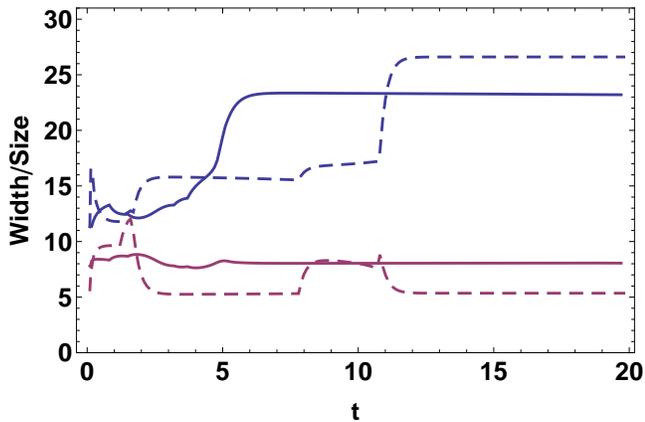}
\caption{
(Color online) The half width of ice structure (upper curves) together with
  the thickness of the boundary (lower curves) without (dashed) and with AFPs (solid) of the time evolution of order parameter from figure \ref{afp_time}.
\label{afp_time1}
}
\end{figure}

In figure \ref{afp_time} we plot the time evolution of an initially small-scale distributed sinusoidal order parameter with and without AFPs. The evolution equations obviously reduce the number of ice grains forming a larger one after some time. Interestingly this accumulation occurs faster with AFPs than without. However, as we have already seen, the absolute height of the ice-order parameter (ideal ice corresponds to $\Psi=1$) is lowered by AFPs during this process. 

This is also illustrated by the time-evolution of the half width of the kinks in figure \ref{afp_time1} which we interpret as the size of the ice grains. One sees that the grain size of ice evolves faster with AFPs and remains at a smaller value compared to the case without AFPs. This means the nucleation of ice starts earlier with AFPs but remains locked at an intermediate stage. This is in agreement with the static observation above that the AFPs support smaller nuclei sizes and inhibit the formation of large clusters at the freezing point. We consider this later blocking of larger cluster sizes as an dynamic process due to the kinetics and coupling of AFPs to the ice embryos described here by a linear term in the free energy. 
Accompanying this observation we also see that the width of the boundary between ice and water remains larger with AFPs than without. This is of course an expression of the reduction of interfacial energy.

\section{Summary}

The interaction of AFP molecules with ice crystals is described by a coupled phase field equation between the order parameter describing ice and water and the AFP concentration. We find essentially two effects of AFPs in suppressing the formation of ice crystals. First the interfacial energy is lowered which allows only smaller ice nuclei to be formed. And secondly we see that the ice grains are 
formed faster by the action of AFPs but become locked at smaller sizes and smaller order parameters. The latter means that the freezing is stopped and  ice-water mixture remains instead of completely freezing. According to the proposed model the AFPs 
do not prevent crystal nucleation, but inhibit further growth 
of the initial crystal nuclei. This essentially dynamic process between AFP structure and ice-order parameter establishes a new possible mechanism for the phenomenon of anti-freeze proteins. We demonstrate that the model is capable to reproduce the experimental data on the freezing point depression in principle. The used phenomenological parameters can be linked to known properties of water \cite{Thoms} though a more microscopic determination of these parameters is highly desired but out of the scope of this paper.

\acknowledgements
We like to thank Gerhard Dieckmann for helpful comments.
This work was supported by DFG-priority program SFB 1158. The financial support by the Brazilian Ministry of Science 
and Technology is acknowledged.






\bibliography{seaice,afp}

\end{document}